\documentclass[aps,prc,superscriptaddress,showpacs,floatfix,nofootinbib,notitlepage,twocolumn]{revtex4-1}
\usepackage{amsmath,graphicx,float,hyperref,upgreek}

\newcommand{\trento}{T$\mathrel{\protect\raisebox{-2.1pt}{R}}$ENTo}

\begin{document}

\title{New paradigm for fluctuations in heavy-ion collisions}

\author{Giuliano Giacalone}
\affiliation{Institut de physique th\'eorique, Universit\'e Paris Saclay, CNRS, CEA, F-91191 Gif-sur-Yvette, France}
\author{Pablo Guerrero-Rodr\'iguez}
\affiliation{CAFPE and Departamento de F\'isica Te\'orica y del Cosmos, Universidad de Granada, E-18071 Campus de Fuentenueva, Granada, Spain}
\affiliation{CPHT, CNRS, Institut Polytechnique de Paris, France}
\author{Matthew Luzum}
\affiliation{Instituto de F\'{i}sica, Universidade de S\~{a}o Paulo, C.P.
66318, 05315-970 S\~{a}o Paulo, Brazil}
\author{Cyrille Marquet}
\affiliation{CPHT, CNRS, Institut Polytechnique de Paris, France}
\author{Jean-Yves Ollitrault}
\affiliation{Institut de physique th\'eorique, Universit\'e Paris Saclay, CNRS, CEA, F-91191 Gif-sur-Yvette, France}

\begin{abstract}
  Since their discovery, fluctuations in the initial state of heavy-ion collisions have been understood as originating mostly from the random positions of nucleons within the colliding nuclei.
  We consider an alternative approach where all the focus is on fluctuations generated by QCD interactions, that we evaluate at leading logarithmic accuracy in the color glass condensate effective theory.
  We validate our approach using BNL Relativistic Heavy Ion Collider (RHIC) and CERN Large Hadron Collider (LHC) data on anisotropic flow. In particular, we show that, compared to standard Glauber-inspired calculations, our formalism provides a better description of the centrality dependence of the ratio of elliptic flow and triangular flow. It also naturally explains the evolution of elliptic flow fluctuations between RHIC and LHC energies.
\end{abstract}

\maketitle

\section{Introduction}

\label{sec:1}

Relativistic heavy-ion collisions are performed at the BNL Relativistic Heavy Ion Collider (RHIC) and at the CERN Large Hadron Collider (LHC) with the aim of creating the quark-gluon plasma, the high-temperature state of strongly-interacting matter.
The nuclear-sized droplet of quark-gluon plasma formed in a collision expands like a low-viscosity fluid~\cite{Heinz:2013th}, whose properties are studied through characteristic azimuthal anisotropies generated during the expansion~\cite{Adams:2004bi,Adler:2003kt,Aamodt:2010pa,ATLAS:2012at,Chatrchyan:2012ta}.
In a hydrodynamic framework, azimuthal anisotropy in the final state is engendered by the spatial anisotropy that characterizes the energy-density profile at the onset of the hydrodynamic evolution~\cite{Heinz:2013th,Yan:2017ivm}.
This \textit{primordial} spatial anisotropy has, in a heavy-ion collision, a twofold origin: 
First, it is due to the almond shape of the overlap area between two nuclei for noncentral collisions, that generates elliptic flow~\cite{Ollitrault:1992bk}; 
 Second, it originates from event-to-event density fluctuations~\cite{Miller:2003kd}, that yield an elliptic deformation even in central collisions~\cite{Alver:2006wh}, and a triangular anisotropy~\cite{Alver:2010gr}.
The role of primordial fluctuations for heavy-ion phenomenology draws, hence, an interesting parallel~\cite{Mishra:2007tw} with the physics of primordial fluctuations in cosmology, where the observed anisotropies of the Cosmic Microwave Background~\cite{Ade:2013kta} originate from quantum fluctuations in the early Universe~\cite{Bardeen:1983qw}.

In the standard picture of heavy-ion collisions, primordial fluctuations originate from the randomness in the spatial positions of the nucleons that populate the wavefunctions of the colliding nuclei~\cite{Miller:2007ri}, with additional contributions at the level of the subnucleonic structure~\cite{Eremin:2003qn,Bialas:2006kw,Loizides:2016djv}.
There may also be fluctuations due to the collision process itself, i.e., to the gluon dynamics~\cite{Lappi:2003bi}.
Putting all these effects together typically results in complex, fully numerical descriptions of the initial state~\cite{ALbacete:2010ad,Schenke:2012wb,Niemi:2015qia} that do not offer an intuitive grasp of the relevant scales and phenomena.

In this paper, we achieve a more transparent description of initial-state fluctuations by applying a recent analytical calculation of energy-density fluctuations~\cite{Albacete:2018bbv} to the phenomenology of anisotropic flow in nucleus-nucleus collisions.
Denoting by $\rho({\bf s})$, where ${\bf s}$ labels a point in the transverse plane, the energy density deposited at mid-rapidity right after a collision takes place, we write that $\rho({\bf s})=\langle \rho({\bf s})\rangle+\delta\rho({\bf s})$, where $\langle \rho({\bf s})\rangle $ is the energy density averaged over many events at a given impact parameter, and $\delta\rho({\bf s})$ is referred to as the fluctuation.
Doing so, the magnitude of density fluctuations, which is given by their variance, or connected two-point function, is:
 \begin{eqnarray}
 \label{defS}
 S({\bf s}_1,{\bf s}_2)  &\equiv& \langle\delta\rho({\bf s}_1)\delta\rho({\bf s}_2)\rangle\cr
 &=&\langle\rho({\bf s}_1)\rho({\bf s}_2)\rangle-\langle\rho({\bf s}_1)\rangle\langle\rho({\bf s}_2)\rangle.
\end{eqnarray}
 Albacete {\it et al.\/}~\cite{Albacete:2018bbv} have calculated $S({\bf s}_1,{\bf s}_2)$ in the color glass condensate~\cite{Iancu:2000hn,Gelis:2010nm,Albacete:2014fwa} (CGC) effective field theory of QCD.
They have expressed $S({\bf s}_1,{\bf s}_2)$ analytically as a function of the saturation scales of the two nuclei and of the relative transverse distance, $r\equiv|{\bf s}_1-{\bf s}_2|$. 
We use this expression as an input to evaluate initial-state anisotropies (Secs.~\ref{sec:2} to~\ref{sec:4}), that we subsequently compare to RHIC and LHC data on anisotropic flow (Sec.~\ref{sec:5}).

\section{Initial-state anisotropies from the 2-point function}

\label{sec:2}

The relevant quantities for phenomenology are dimensionless complex Fourier coefficients that characterize the spatial anisotropy of the initial density field, $\rho({\bf s})$. 
They are defined, in a centered coordinate system,\footnote{We mean that the center of energy lies at the origin, $\int_{\bf s}{\bf s}\,\rho({\bf s})=0$.} as~\cite{Teaney:2010vd,Qiu:2011iv}
\begin{equation}
\label{defepsn}
\varepsilon_n \equiv \frac{\int_{{\bf s}} \, {\bf s}^n \rho({\bf s})}{\int_{{\bf s}} \, |{\bf s}|^n\rho({\bf s})},
\end{equation}
where we use the complex coordinate ${\bf s}=x+iy$, and the short hand $\int_{{\bf s}}=\int {\rm d}x{\rm d}y$ for the integration over the transverse  plane. 
$\varepsilon_2$ and $\varepsilon_3$ thus defined quantify, respectively, the amount of elliptic and triangular deformation of the density profile.

The coefficient of anisotropic flow, $v_n$, is defined as the $n$-th Fourier harmonic of the azimuthal distribution of outgoing particles~\cite{Luzum:2011mm}. 
The largest harmonics in the spectrum are elliptic flow, $v_2$, and triangular flow, $v_3$. 
Hydrodynamic simulations show that $v_n$ is to a good approximation linearly correlated with $\varepsilon_n$ in a narrow bin of centrality~\cite{Gardim:2011xv,Niemi:2012aj,Noronha-Hostler:2015dbi}, so that to a first approximation one can simply write $v_n = \kappa_n \varepsilon_n$ on a event-by-event basis. The response coefficient, $\kappa_n$, depends very mildly on the impact parameter of the collision, for a given colliding system and energy~\cite{Noronha-Hostler:2015dbi}.

Anisotropic flow, though, is not measured on an event-by-event basis, but inferred from correlations which are averaged over events in a given class of collision centrality. The default measure of $v_n$ is an rms average, denoted by $v_n\{2\}\equiv\langle |v_n|^2\rangle^{1/2}$~\cite{Luzum:2012da}, where the 2 inside curly brackets means that it is inferred from analyses of 2-particle correlations. 
Linear hydrodynamic response, then, implies $v_n\{2\}=\kappa_n\varepsilon_n\{2\}$.
Thus, the relevant quantity coming from the initial state is the rms average of $\varepsilon_n$, denoted by $\varepsilon_n\{2\}$.
It turns out that this quantity can be expressed in terms of the two-point function, $S({\bf s}_1,{\bf s}_2)$, under minimal assumptions~\cite{Blaizot:2014nia}.

Let us start with the simple case of a collision at zero impact parameter.
Since the mean density profile, $\langle\rho({\bf s})\rangle$, is azimuthally symmetric, one can replace $\rho({\bf s})$ with $\delta\rho({\bf s})$ in the numerator of Eq.~(\ref{defepsn}).
To leading order in the fluctuation, $\delta\rho({\bf s})$, then, one can replace 
$\rho({\bf s})$ with $\langle\rho({\bf s})\rangle$ in the denominator.
Multiplying by the complex conjugate, $\varepsilon_n^*$, and averaging over events, one immediately obtains~\cite{Blaizot:2014nia}:
\begin{equation}
\label{epBcentral}
  \varepsilon_n\{2\}^2\equiv\langle|\varepsilon_n|^2\rangle=\frac{\int_{{\bf s}_1,{\bf s}_2} \,({\bf s}_1)^n\,({\bf s}_2^*)^n\, S({\bf s}_1,{\bf s}_2)}{\left( \int_{{\bf s}}\, |{\bf s}|^n\langle\rho({\bf s}) \rangle  \right)^2}.
\end{equation}
We now generalize to non-central collisions. The main difference is that the mean density profile, $\langle\rho({\bf s})\rangle$, is no longer isotropic, but has an elliptic shape.
Its departure from isotropy is quantified by the mean anisotropy $\bar\varepsilon_2$, which is given by replacing $\rho({\bf s})$ with $\langle\rho({\bf s})\rangle$ in Eq.~(\ref{defepsn}):
\begin{equation}
\label{defepBbar}
  \bar\varepsilon_2\equiv \frac{\int_{{\bf s}} \, {\bf s}^2 \langle\rho({\bf s})\rangle}{\int_{{\bf s}} \, |{\bf s}|^2\langle\rho({\bf s})\rangle}.
\end{equation}
This is a quantity of direct phenomenological relevance.
Indeed, the fourth-cumulant measure of elliptic flow, $v_2\{4\}\equiv\left(2\langle v_2^2\rangle^2-\langle v_n^4\rangle\right)^{1/4}$~\cite{Borghini:2001vi}, is, in the regime of linear hydrodynamic response, equal to $v_2\{4\}=\kappa_2\varepsilon_2\{4\}$, where $\varepsilon_2\{4\}$ is the fourth-order cumulant of $\varepsilon_2$ fluctuations, that can be taken as:
\begin{equation}
\label{epB4}
\varepsilon_2\{4\}\approx\bar\varepsilon_2.
\end{equation}
Equation~(\ref{epB4}) assumes that $\bar\varepsilon_2$ coincides with the mean eccentricity in the reaction plane~\cite{Bhalerao:2006tp} and that eccentricity fluctuations are Gaussian~\cite{Voloshin:2007pc,Floerchinger:2014fta}.  
This turns out to be a very good approximation for collisions up to $\sim 30\%$ centrality, beyond which non-Gaussian corrections become sizable~\cite{Giacalone:2016eyu,Sirunyan:2017fts,Acharya:2018lmh,Mehrabpour:2018kjs,Bhalerao:2018anl}.

The total rms eccentricity, $\varepsilon_2\{2\}$, is obtained by adding in quadrature the mean eccentricity and the contribution of fluctuations, which is the right-hand side of Eq.~(\ref{epBcentral}).
Therefore, for non-central collisions, we simply replace Eq.~(\ref{epBcentral}) with 
\begin{eqnarray}
  \label{epBnoncentral}
\sigma^2\equiv\varepsilon_2\{2\}^2-\bar\varepsilon_2^2&=&
\frac{\int_{{\bf s}_1,{\bf s}_2} \,({\bf s}_1)^2\,({\bf s}_2^*)^2\, S({\bf s}_1,{\bf s}_2)}{\left( \int_{{\bf s}}\, |{\bf s}|^2\langle\rho({\bf s}) \rangle  \right)^2}\cr
  \varepsilon_3\{2\}^2&=&\frac{\int_{{\bf s}_1,{\bf s}_2} \,({\bf s}_1)^3\,({\bf s}_2^*)^3\, S({\bf s}_1,{\bf s}_2)}{\left( \int_{{\bf s}}\, |{\bf s}|^3\langle\rho({\bf s}) \rangle  \right)^2},
\end{eqnarray}
where we introduce the notation $\sigma^2$ for the variance of $\varepsilon_2$ fluctuations.
A more careful treatment of non-central collisions is carried out in Ref.~\cite{Bhalerao:2019uzw}, and yields more complicated expressions.
However, the changes in the results are numerically small, so that the above equations constitute good approximations in practice.

\section{1- and 2-point functions from the CGC}

\label{sec:3}

Derivations of the initial average energy density, $\langle\rho({\bf s})\rangle$, in the CGC framework date back to several years~\cite{Lappi:2006hq,Chen:2015wia}.
Following Ref.~\cite{Albacete:2018bbv}, with $N_c=3$, it simply reads:
\begin{equation}
\label{eq:1p}
\langle\rho({\bf s})\rangle = \frac{4}{3g^2} Q_{A}^2({\bf s}) Q_{B}^2({\bf s}),
\end{equation}
where subscripts $A$ and $B$ label the two colliding nuclei, $g$ is the strong coupling constant, and $Q_{A,B}({\bf s})$ is the local saturation scale of the nucleus.
$Q_{A}^2({\bf s})$ is proportional to the density of nucleons per transverse area at point ${\bf s}$, which is traditionally denoted by $T_A({\bf s})$, and is obtained by integrating the nuclear density over the longitudinal coordinate~\cite{Miller:2007ri}.\footnote{More explicitly, we use $Q_s^2({\bf s})=Q_{s0}^2 T({\bf s})/T({\bf 0})$, where $Q_{s0}$ is the value of the saturation scale at the center of the nucleus.}
Injecting Eq.~(\ref{eq:1p}) into Eq.~(\ref{defepBbar}), and to the extent that the nuclear density is known, one obtains a parameter-free prediction for the average eccentricity of the system~\cite{Lappi:2006xc}:
\begin{equation}
\label{epBbarcgc}
\bar\varepsilon_2\equiv \frac{\int_{{\bf s}} \, {\bf s}^2 T_A({\bf s})T_B({\bf s})}{\int_{{\bf s}} \, |{\bf s}|^2 T_A({\bf s})T_B({\bf s})}.
\end{equation}

The crucial new information coming from the CGC theory is the connected two-point function, $S({\bf s}_1,{\bf s}_2)$, computed in Ref.~\cite{Albacete:2018bbv}, which allows us to evaluate $\varepsilon_2\{2\}$ and $\varepsilon_3\{2\}$, as given by Eq.~(\ref{epBnoncentral}).
The CGC typically predicts that the energy-density fluctuations are correlated over a transverse extent of order $1/Q_s$~\cite{Lappi:2015vha}, which is much shorter than the nuclear radius, $R$. 
In other terms, $S({\bf s}_1,{\bf s}_2)$ is small if $r=|{\bf s}_1-{\bf s}_2|\gg 1/Q_s$.
Therefore, it is natural to change variables to ${\bf s}_1={\bf s}+{\bf r}/2$, ${\bf s}_2={\bf s}-{\bf r}/2$, and integrate first over ${\bf r}$.
Albacete {\it et al.\/} show that $S( {\bf s}+{\bf r}/2,{\bf s}-{\bf r}/2)$ falls off slowly at large distances, like $1/r^2$, so that its integral is logarithmically divergent. 
It must be regulated by an infrared cutoff, dubbed $m$.

Hence, imposing the following separation of scales:
\begin{equation}
\label{eq:separation}
  \frac{1}{Q_s}\ll\frac{1}{m}\ll R,
\end{equation}
where $R$ is the transverse size of the system, one can take $S({\bf s}_1,{\bf s}_2)$ from Ref.~\cite{Albacete:2018bbv}, whose integral over ${\bf r}$ yields, to leading logarithmic accuracy:
\begin{widetext}
\begin{equation}
\label{eq:2p}
  \xi({\bf s})\equiv
  \int_{\bf r}S\left( {\bf s}+\frac{{\bf r}}{2},{\bf s}-\frac{{\bf r}}{2}\right)
  =\frac{16\pi}{9 g^4}  Q_{A}^2({\bf s}) Q_{B}^2({\bf s})
  \left[Q_{A}^2({\bf s})\ln\left(1+\frac{Q_{B}^2({\bf s})}{m^2}\right)
  +Q_{B}^2({\bf s})\ln\left(1+\frac{Q_{A}^2({\bf s})}{m^2}\right)\right].
\end{equation} 
\end{widetext}
The derivation of Eq.~(\ref{eq:2p}) is detailed in Appendix~\ref{sec:a}.

Assuming that the range of correlation is much smaller than the nuclear radius, Equations~(\ref{epBnoncentral}) give~\cite{Blaizot:2014nia}:
\begin{eqnarray}
\label{eq:epBfinal}
\sigma^2=\varepsilon_2\{2\}^2-\bar\varepsilon_2^2&=&
\frac{\int_{\bf s} \,|{\bf s}|^4\,\xi({\bf s})}{\left( \int_{{\bf s}}\, |{\bf s}|^2\langle\rho({\bf s}) \rangle  \right)^2}\cr
\varepsilon_3\{2\}^2&=&\frac{\int_{\bf s} \,|{\bf s}|^6\,\xi({\bf s})}{\left( \int_{{\bf s}}\, |{\bf s}|^3\langle\rho({\bf s}) \rangle  \right)^2}.
\end{eqnarray}
These equations express the variance of anisotropy coefficients as a function of $\xi({\bf s})$, which represents the density of variance of the initial density field~\cite{Bhalerao:2019uzw}. 

\section{Evaluating initial-state anisotropy}
\label{sec:4}

We now present quantitative results for initial anisotropies.
Inserting Eqs.~(\ref{eq:1p}) and (\ref{eq:2p}) into Eqs.~(\ref{eq:epBfinal}), one sees that the coupling constant, $g$, cancels between the numerator and the denominator.
There are two free parameters in our calculation:
The most important is the proportionality constant between $(Q_A)^2$ and $T_A$, or, equivalently, the value of the saturation scale $Q_A$ at the center of the nucleus, which we denote by $Q_{s0}$. 
The other free parameter is the infrared cutoff $m$.

We start by evaluating orders of magnitude. 
In the simple case of a central collision ($b=0$, which implies $\bar\varepsilon_2=0$) and a uniform density profile within a disk of radius $R$, Eqs.~(\ref{eq:epBfinal}) give:
\begin{eqnarray}
\label{eq:control}
\varepsilon_2\{2\}^2&=&\frac{8}{3}\frac{1}{(Q_{s0}R)^2}\ln\left(\frac{Q_{s0}^2}{m^2}\right)
\cr
\varepsilon_3\{2\}^2&=&\frac{25}{8}\frac{1}{(Q_{s0}R)^2}\ln\left(\frac{Q_{s0}^2}{m^2}\right).
\end{eqnarray}
The traditional picture is that the initial state is made of independent color domains of transverse size $\sim 1/Q_{s0}$~\cite{Lappi:2015vha}. The number of independent domains is $N\sim (Q_{s0}R)^2$, and the variance of $\varepsilon_n$ fluctuations is of order $1/N$~\cite{Bhalerao:2006tp,Bhalerao:2011bp}. 
Note, however, that fluctuations are enhanced by a large logarithm in Eq.~(\ref{eq:control}), which is due to the slow fall-off of the correlation function at large distances (see Appendix~\ref{sec:a}). 
For $Q_{s0}=1$~GeV, $m=0.14$~GeV, $R=6.5$~fm, Eq.~(\ref{eq:control}) gives $\varepsilon_2\{2\}\simeq\varepsilon_3\{2\}\simeq 0.1$. 


Eq.~(\ref{eq:control}) also allows us to assess the effect of varying the infrared cutoff $m$.
One expects results to be essentially independent of $m$ provided that one renormalizes $Q_{s0}$ in such a way that the right-hand side of Eq.~(\ref{eq:control}) is unchanged.
We note that, in turn, our approach does not allow to determine $m$ and $Q_{s0}$ separately.
This could be achieved using observables that probe the short-range structures of the initial density profile. 
Arguably, this is not feasible with the observables considered here, i.e., flow observables in nucleus-nucleus systems, that are remarkably insensitive to density fluctuations over subnucleonic scales~\cite{Gardim:2017ruc}.
The effect of varying $m$, and the resulting shift of $Q_{s0}$, are studied in detail in Appendix~\ref{sec:b}.

\begin{figure*}[t!]
    \centering
    \includegraphics[width=.75\linewidth]{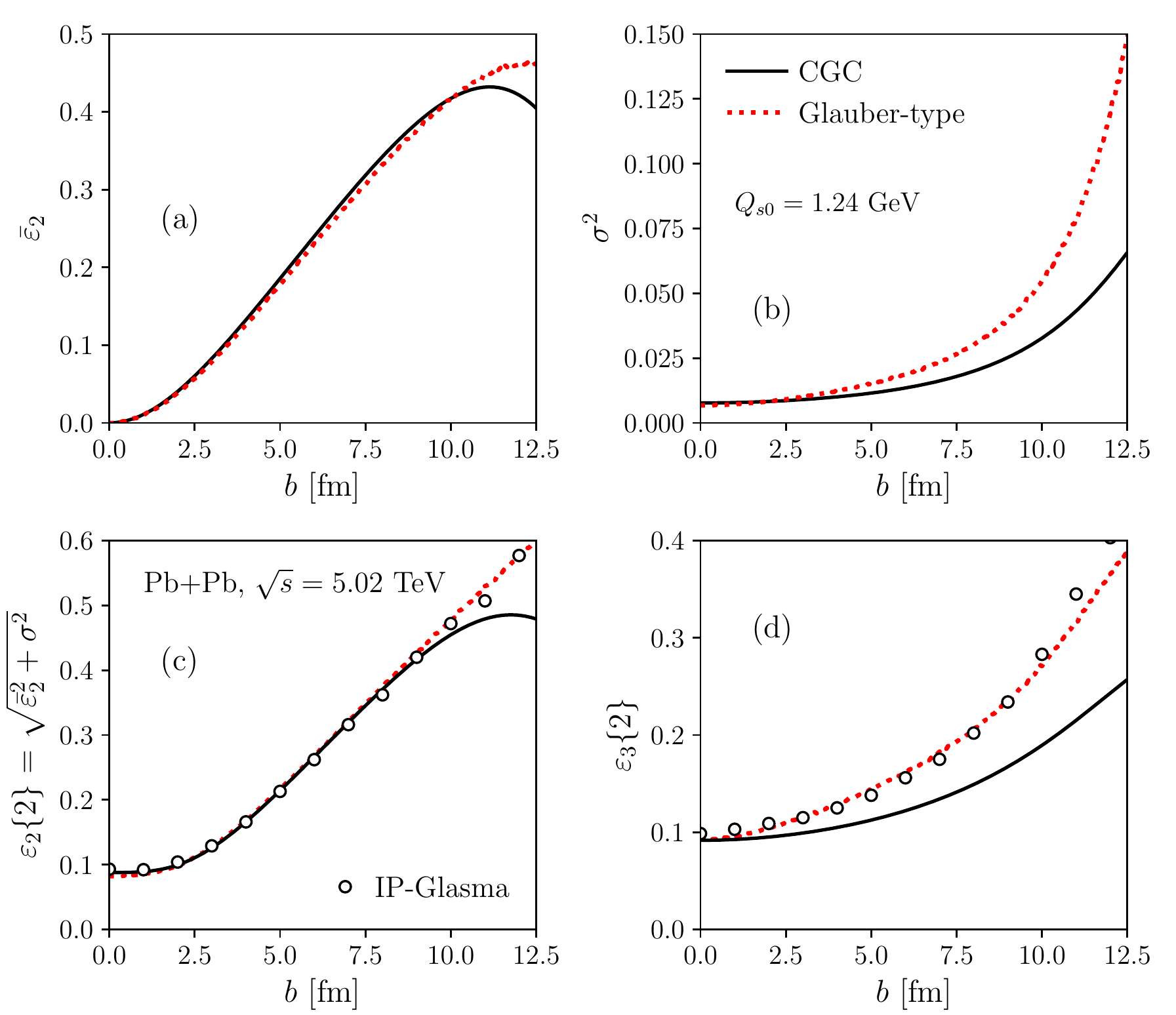}
    \caption{Anisotropy fluctuations as function of impact parameter in 5.02 TeV Pb+Pb collisions. 
      The solid line in panel (a) corresponds to Eq.~(\ref{epBbarcgc}), while in panels (b) and (d) it corresponds to Eqs.~(\ref{eq:epBfinal}). The calculation is performed for $Q_{s0}=1.24$ GeV and $m=0.14$ GeV.
      Dotted lines correspond to the \trento{} model tuned to LHC data~\cite{Moreland:2014oya}.
Symbols in panels (c) and (d) are results from the IP-Glasma model~\cite{Schenke:2012wb}.}
    \label{fig:1}
\end{figure*}

We now present quantitative results for initial anisotropies in collisions of $^{208}$Pb nuclei at $\sqrt{s_{\rm NN}}=5.02$~TeV, as a function of impact parameter. 
We take the nuclear matter density as a 2-parameter Fermi distribution, with parameters from Ref.~\cite{DeJager:1987qc}. 
It is natural to ask how results obtained within our CGC formalism compare with those of state-of-the-art Monte Carlo models of nucleus-nucleus collisions.
To this purpose, we shall compare our results to the eccentricity harmonics provided by the \trento{} model~\cite{Moreland:2014oya}.
Note that, while comparing our results to \trento{}, we are effectively testing our calculations against a wide range of initial-state models for heavy-ion collisions.
Glauber-inspired Monte Carlo models of initial conditions are known to present roughly the same $\varepsilon_3$, and the same $\varepsilon_2$ close to $b=0$ (see Refs.~\cite{Moreland:2014oya,Nagle:2018ybc} for exhaustive comparisons).
In view of this, in our figures we shall refer to the \trento{} calculation as a ``Glauber-type'' calculation.

To begin with, we evaluate the average anisotropy of the system, given by Eq.~(\ref{epBbarcgc}).
CGC and \trento{} give essentially the same results, as shown in Fig.~\ref{fig:1}(a).
Less trivial is the evaluation of Eqs.~(\ref{eq:epBfinal}), that allows us to study the difference between our new paradigm for fluctuation and the standard one based on fluctuating positions of nucleons.
Results from \trento{} are shown as dotted lines in Fig.~\ref{fig:1} (b), (c), (d). 
Fluctuations from Eqs.~(\ref{eq:epBfinal}) are shown as solid lines, for $Q_{s0}=1.24$~GeV and $m=0.14$ GeV.
We note that, with this choice of the parameters, CGC and \trento{} give similar results.
For both models, fluctuations (as measured by $\sigma^2$ and $\varepsilon_3\{2\}$ in panels (b) and (d)) increase as a function of impact parameter, which is understood as a natural consequence of the smaller system size.
A closer examination reveals that the increase is milder in our CGC calculation.
This feature turns out to be crucial for phenomenological applications, as we shall discuss in Sec.~\ref{sec:5}.

Finally, for completeness we show as well results from the IP-Glasma model~\cite{Schenke:2012wb} in Fig.~\ref{fig:1} (c), (d). 
As expected from previous comparisons~\cite{Schenke:2012fw,Nagle:2018ybc}, IP-Glasma results turn out to be essentially identical to those of the \trento{} model, and therefore differ somewhat from our Glasma calculation.

Note that the CGC framework uses expressions for $\langle\rho({\bf s})\rangle$ and $S({\bf s}_1,{\bf s}_2)$ that describe the system right after the collision takes place, whereas \trento{} gives the entropy profile of the system at the beginning of hydrodynamics.
This difference is not important, as classical Yang-Mills evolution to a finite proper time does not modify the values of $\varepsilon_n$ \cite{Schenke:2012fw}.

\begin{figure*}[t!]
    \centering
    \includegraphics[width=\linewidth]{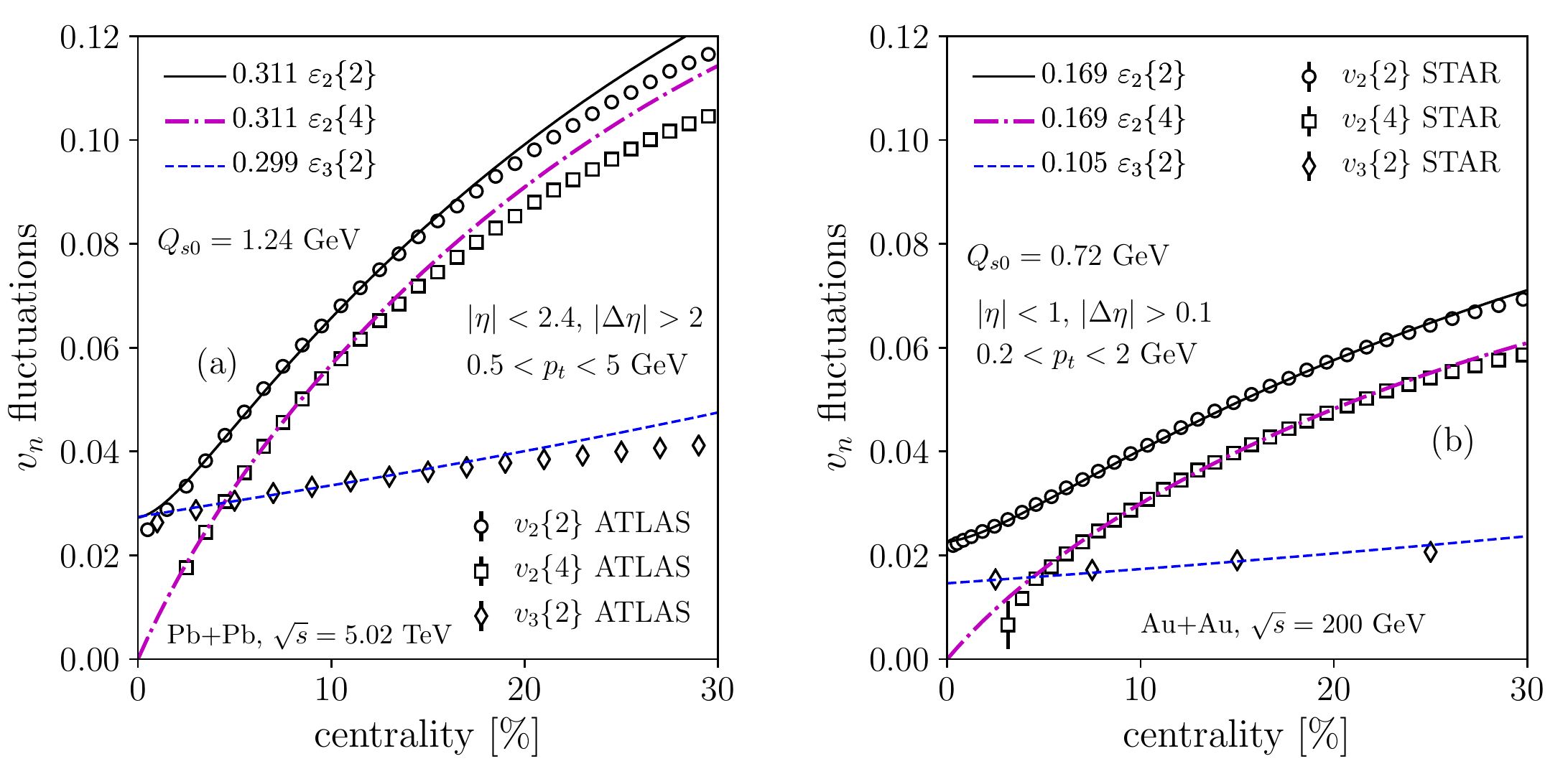}
    \caption{Symbols: Experimental data on $v_2$ and $v_3$, as function of centrality percentile, measured by the ATLAS Collaboration~\cite{Aaboud:2019sma} in 5.02 TeV Pb+Pb collisions [panel (a)], and by the STAR Collaboration~\cite{Adamczyk:2015obl} in 200 GeV Au+Au collisions [panel (b)]. Lines represent results from our CGC formalism, rescaled according to Eq.~(\ref{rescaling}) for $m=0.14$ GeV and best-fit values of $Q_{s0}$, depending on the collision energy. The extracted values of $\kappa_2$ and $\kappa_3$ at both RHIC and LHC energy are displayed in the legends as factors multiplying the cumulants of $\varepsilon_n$ fluctuations.}
    \label{fig:2}
\end{figure*}

\section{Comparison with RHIC and LHC data}
\label{sec:5}

We now compare our CGC calculations to experimental data on $v_2\{2\}$, $v_2\{4\}$ and $v_3\{2\}$ in Pb+Pb collisions at ~$\sqrt{s_{\rm NN}}=5.02$~TeV~\cite{Aaboud:2019sma}, and in Au+Au collisions at ~$\sqrt{s_{\rm NN}}=200$~GeV~\cite{Adamczyk:2015obl}.
We restrict our theory-to-data comparison to the 0-30\% centrality range, in which $\kappa_n$ is essentially constant in hydrodynamics~\cite{Noronha-Hostler:2015dbi}.
We use the geometric relation between the impact parameter and the centrality of a collision to express our results as function of the centrality percentile,\footnote{Note that in experiment, the centrality of the collision is defined in a different way. However, the geometric definition is a very good approximation in practice, except for the most central collisions~\cite{Das:2017ned}.} 
i.e., we use $centrality = (\pi b^2)/\sigma_{\rm inel}$,
where $\sigma_{\rm inel}$ is the inelastic nucleus-nucleus cross section, which we take from the Glauber model: $\sigma_{\rm inel}=685$ fm$^2$ for Au+Au collisions, and $\sigma_{\rm inel}=767$ fm$^2$ for Pb+Pb collisions.
Linear hydrodynamic response implies the following relations between final-state flow harmonics and initial-state anisotropies: 
\begin{eqnarray}
  \label{rescaling}
v_2\{2\}&=&\kappa_2\varepsilon_2\{2\},\cr
v_2\{4\}&=&\kappa_2\bar\varepsilon_2,\cr
v_3\{2\}&=&\kappa_3\varepsilon_3\{2\}.
\end{eqnarray}
We treat $\kappa_2$, $\kappa_3$, and $Q_{s0}$ as free parameters, which we adjust to data.
In the following results, the value of $m$ is always chosen equal to $0.14$ GeV.

\subsection{Cumulants of flow fluctuations}

Since the mean eccentricity in the reaction plane, $\bar\varepsilon_2$, in Eq.~(\ref{epBbarcgc}) does not depend on $Q_{s0}$, we first use $v_2\{4\}$ to fix the value of $\kappa_2$.
The dot-dashed lines in Fig.~\ref{fig:2} show that our calculation captures the measured centrality dependence of $v_2\{4\}$, both at RHIC and at LHC.\footnote{The sharp decrease of $v_2\{4\}$ at RHIC below 5\% centrality is an effect of centrality fluctuations~\cite{Zhou:2018fxx}, which are not included in our description.}
We note that our formula leads to a better description of RHIC data, which are essentially captured all the way up to 25\% centrality.
This finding suggests that either elliptic flow fluctuations at LHC energy are in general less Gaussian than at RHIC energy, so that the approximation $\varepsilon_2\{4\}\approx\bar\varepsilon_2$ is less justified at LHC energy, or that the centrality dependence of the response coefficient $\kappa_2$ is stronger at LHC energy, a feature that has never been investigated in hydrodynamic simulations.

With the knowledge of $\kappa_2$ at hand, we move on to the description of $v_2\{2\}=\kappa_2\varepsilon_2\{2\}$.
This quantity is less trivial because it depends on $Q_{s0}$.
The solid lines in Fig.~\ref{fig:2} show the rescaled $\varepsilon_2\{2\}$ corresponding to $Q_{s0}=1.24$~GeV. 
This result is in very good agreement with the measured $v_2\{2\}$.
As expected, we observe a significant energy dependence of $Q_{s0}$: LHC data [panel (a)] are reproduced in our calculation with a larger value of value of $Q_{s0}$, of order 1.24~GeV at LHC energy versus 0.72~GeV at RHIC energy.
We come back to this point below in Sec.~\ref{sec:edep}. 

Finally, we fit the value of $\kappa_3$ to match the value of $v_3\{2\}$ in central collisions.
Our results are displayed as dashed lines in Fig.~\ref{fig:2}.
Agreement with data is very good throughout the chosen centrality range.

\subsection{Triangular flow and the ratio $\boldsymbol{v_2\{2\}/v_3\{2\}}$}

\begin{figure}[t!]
    \centering
    \includegraphics[width=.99\linewidth]{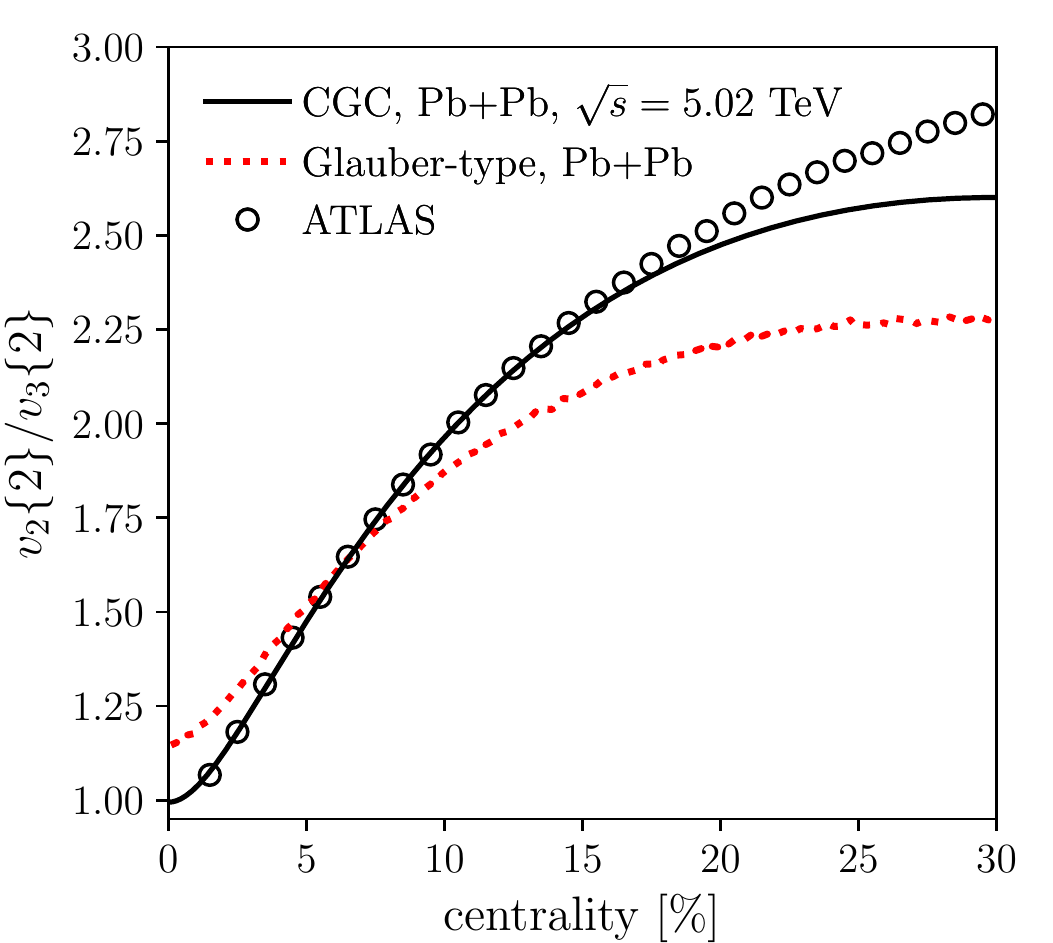}
    \caption{Symbols: Ratio $v_2\{2\}/v_3\{2\}$ from ATLAS data~\cite{Aaboud:2019sma}. Solid line: CGC calculation, taking $\kappa_2$ and $\kappa_3$ from Fig.~\ref{fig:2}. Dotted line: \trento{} model, with $\kappa_2$ and $\kappa_3$ extracted using the same method as in Fig.~\ref{fig:2}.}
    \label{fig:3}
\end{figure}

The centrality dependence of the ratio $v_2\{2\}/v_3\{2\}$ is typically steeper in experiment than in hydrodynamic calculations.
This was shown explicitly in Ref.~\cite{Alba:2017hhe}, but can also be inferred from previous articles~\cite{Shen:2015qta,Rose:2014fba}.
We present our results for this ratio, along with ATLAS data, as a solid line in Fig.~\ref{fig:3}.
Agreement with data is much better than with the \trento{} model, shown as a dotted line.
The \trento{} model predicts a less steep dependence, as found in previous hydrodynamic calculations.
The key feature that restores agreement with data is the mild growth of $\varepsilon_3$ with impact parameter in our CGC calculation (Fig.~\ref{fig:1} (d)). 

\subsection{Relative $\boldsymbol{v_2}$ fluctuations and energy dependence}
\label{sec:edep}

\begin{figure}[t!]
    \centering
    \includegraphics[width=\linewidth]{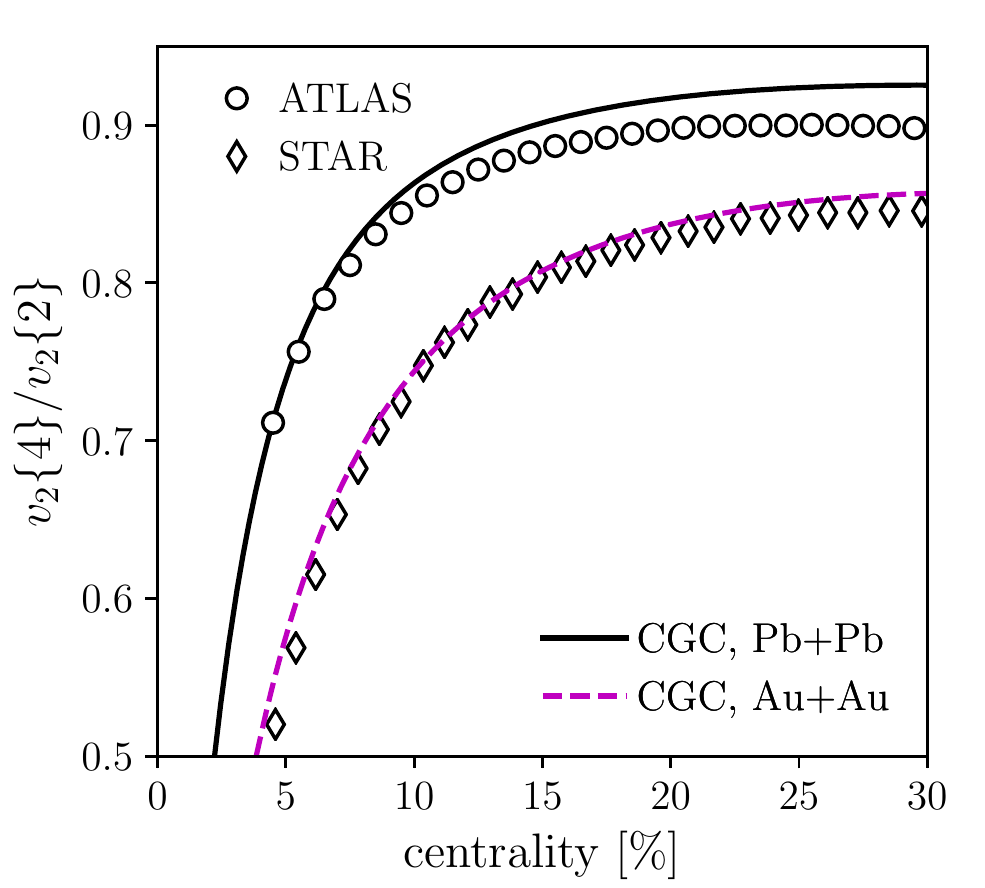}
    \caption{$v_2\{4\}/v_2\{2\}$ as function of centrality percentile. Circles: ATLAS data for Pb+Pb collisions at~ $\sqrt{s}=5.02$~TeV~\cite{Aaboud:2019sma}. Diamonds: STAR data for Au+Au collisions at~ $\sqrt{s}=200$~GeV~\cite{Adamczyk:2015obl}. Solid line: CGC calculation for Pb+Pb collisions, with $Q_{s0}=1.24$~GeV. Dashed line: CGC calculation for Au+Au collisions, with $Q_{s0}=0.72$~GeV. The infrared cutoff is equal to $0.14$~GeV.}
    \label{fig:4}
\end{figure}

The second fluctation-driven observable that is relevant to test our approach is the splitting between $v_2\{2\}$ and $v_2\{4\}$, i.e., the relative fluctuations of elliptic flow.
Relative fluctuations are conveniently quantified using the ratio $v_2\{4\}/v_2\{2\}$, which is equal to $\varepsilon_2\{4\}/\varepsilon_2\{2\}$ in the regime of linear response~\cite{Giacalone:2017uqx}. 
This ratio typically goes to 0 for central collisions where $\varepsilon_2\{4\}$ vanishes, and grows quickly to values close to unity in peripheral collisions, where anisotropy is driven by the geometry of the nuclear overlap.
We show $v_2\{4\}/v_2\{2\}$ measured by both the ATLAS and the STAR Collaborations in Fig.~{\ref{fig:4}}.
We observe a very pronounced difference between these two results, implying that elliptic flow fluctuations are significantly larger at RHIC energy than at LHC energy.

The CGC formalism provides a transparent explanation for this phenomenon.
Following Eq.~(\ref{eq:control}), fluctuations are larger at lower energy, because of the lower $Q_{s0}$.
Therefore, in our picture the larger splitting between $v_2\{2\}$ and $v_2\{4\}$ observed in RHIC data is simply a consequence of the energy dependence of the saturation scale.
Our results from the CGC are displayed as lines in Fig.~\ref{fig:4}.
They provide a very good description of data.

It is crucial to appreciate that our extracted values of $Q_{s0}$ grow by factor close to $1.6$ from RHIC to LHC energy.
Now, fits of deep inelastic electron-proton data indicate that the energy dependence of the saturation scale follows~\cite{Albacete:2014fwa}:
\begin{equation}
\frac{Q_s^2[{\rm LHC}]}{Q_s^2[{\rm RHIC}]} = \biggl(\frac{\sqrt{s_{\rm LHC}}}{\sqrt{s_{\rm RHIC}}}\biggr)^{0.28} = 1.57.
\end{equation}
Our results, then, are consistent with the small-$x$ scaling of $Q_s$ obtained from deep inelastic scattering data, a feature which is usually used as input for the modeling of the initial state.

We further stress that standard Glauber Monte Carlo calculations do not make any specific predictions for the energy dependence of $v_2\{4\}/v_2\{2\}$.
It would be interesting to see $v_2\{4\}/v_2\{2\}$ computed in the IP-Glasma approach~\cite{Schenke:2012wb} at both RHIC and LHC energies.

\section{Discussion and outlook}

\label{sec:6}

We have shown that energy-density fluctuations calculated at leading logarithmic accuracy in the CGC effective theory yield initial-state anisotropies which allow us to match the values of $v_2\{2\}$, $v_2\{4\}$ and $v_3\{2\}$ measured in central to midcentral nucleus-nucleus collisions at the RHIC and at the LHC.
This is obtained for values of $\kappa_2$ and $\kappa_3$ that are reasonable, i.e., they are compatible with those found in state-of-the art viscous hydrodynamic calculations~\cite{Noronha-Hostler:2015dbi}.\footnote{One must take into account the larger $p_t$ cut of the ATLAS analysis, which significantly increases $\kappa_2$ and $\kappa_3$~\cite{Luzum:2010ag}.}

Fluctuations of energy density in our approach are entirely given by the variance of the local number of color sources, or large-$x$ partons, on top of a mean matter density background given by the thickness of the considered nucleus.
This new paradigm frees the modeling of initial-state fluctuations from typical strong assumptions made about the role of the nucleons.
It does not rely on an initial random sampling of positions of nucleons, nor on any prescription about the interaction of nucleons and the subsequent deposition of energy.

Note that our approach differs as well from the IP-Glasma model~\cite{Schenke:2012wb}.
Let us recall the generic definition of the (local) saturation scale:
\begin{equation}
\label{eq:qs}
Q_s({\bf s}) \propto h({\bf s})\int_{-\infty}^{+\infty} \mu^2(s^-) ds^- ,
\end{equation}
where $h({\bf s})$ represents the local transverse density of nuclear matter, and $\mu^2(s^-)$ is the local density of color charges along the coordinate where large-$x$ partons appear as frozen sources of color fields.

In the IP-Glasma approach, the function $h({\bf s})$ fluctuates on an event-by-event basis, because $Q_s({\bf s})$ is evaluated at the level of the individual nucleons, whose positions are sampled randomly in each nuclear configuration.
In IP-Glasma, then, the variance of $\mu^2(s^-)$ represents typically a small contribution to the large event-by-event fluctuations of $h({\bf s})$, as also indicated by the curves shown in Fig.~\ref{fig:1}.
In our approach, we keep $h({\bf s})$ as a fixed mean density background, so that all fluctuations are given by the variance of $\mu^2(s^-)$.
The statement that fluctuations of energy density are larger at RHIC energy because of the lower saturation scale, in our formalism simply means that the variance of $\mu^2(s^-)$ is larger at RHIC energy because there are fewer color charges. 

Note that, since color charge fluctuations in our approach compensate for other sources of fluctuations, the values of $Q_{s0}$ that we extract from data are smaller than typical values of $Q_s$ used in the literature, although they are very reasonable~\cite{Kowalski:2007rw,Dusling:2009ni}.
It would be interesting to perform further tests of our paradigm using observables that probe fluctuations in the nuclear wavefunctions, and that can be studied through deep inelastic scattering of electrons on nuclei (see, e.g., Refs.~\cite{Kowalski:2008sa, Mantysaari:2017slo}). 
This could bring new insight on observables that will be investigated at the future electron-ion collider.

Finally, we remark that the system created by the interaction of two CGCs does not boil down to an ideal gas of identical pointlike sources.
The statistics of energy fluctuations in an ideal gas follows Poisson statistics, with a variance proportional to the mean. 
By contrast, in the CGC picture the mean is proportional to $Q_s^4$ [Eq.~(\ref{eq:1p})], while the variance is proportional to $Q_s^6$ [Eq.~(\ref{eq:2p})], neglecting the smoothly-varying logarithm. 
This enhances the role of fluctuations at the center of the fireball, where the system is denser.
The consequences of this nontrivial prediction of high-energy QCD, which follows essentially from dimensional analysis, deserve further investigations.

\section{acknowledgements}
J-Y.O. thanks Jean-Paul Blaizot, Wojciech Broniowski, Fran\c cois Gelis, Dong Jo Kim, Tuomas Lappi, Aleksas Mazeliauskas for discussions.
G.G. acknowledges Stephan Floerchinger, Eduardo Grossi, Heikki M\"antysaari, Aleksas Mazeliauskas, Jan Pawlowski, Raju Venugopalan for stimulating discussions. 
We thank Bjoern Schenke and Raju Venugopalan for useful comments on the first version of this manuscript.
We thank Jacquelyn Noronha-Hostler for suggesting us to study the ratio $v_2\{2\}/v_3\{2\}$.
M.L.~acknowledges support from FAPESP projects 2016/24029-6  and 2017/05685-2, and project INCT-FNA Proc.~No.~464898/2014-5.
M.L. and G.G. acknowledge funding from the USP-COFECUB project Uc Ph 160-16 (2015/13).
P.G-R. acknowledges financial support from the `La Caixa' Banking Foundation.
The work of CM was supported in part by the Agence Nationale de la Recherche under the project ANR-16-CE31-0019-02. 

\appendix
\section{Density of variance at leading logarithmic accuracy in the MV model}
\label{sec:a}

\begin{figure}[b!]
    \centering
    \includegraphics[width=.8\linewidth]{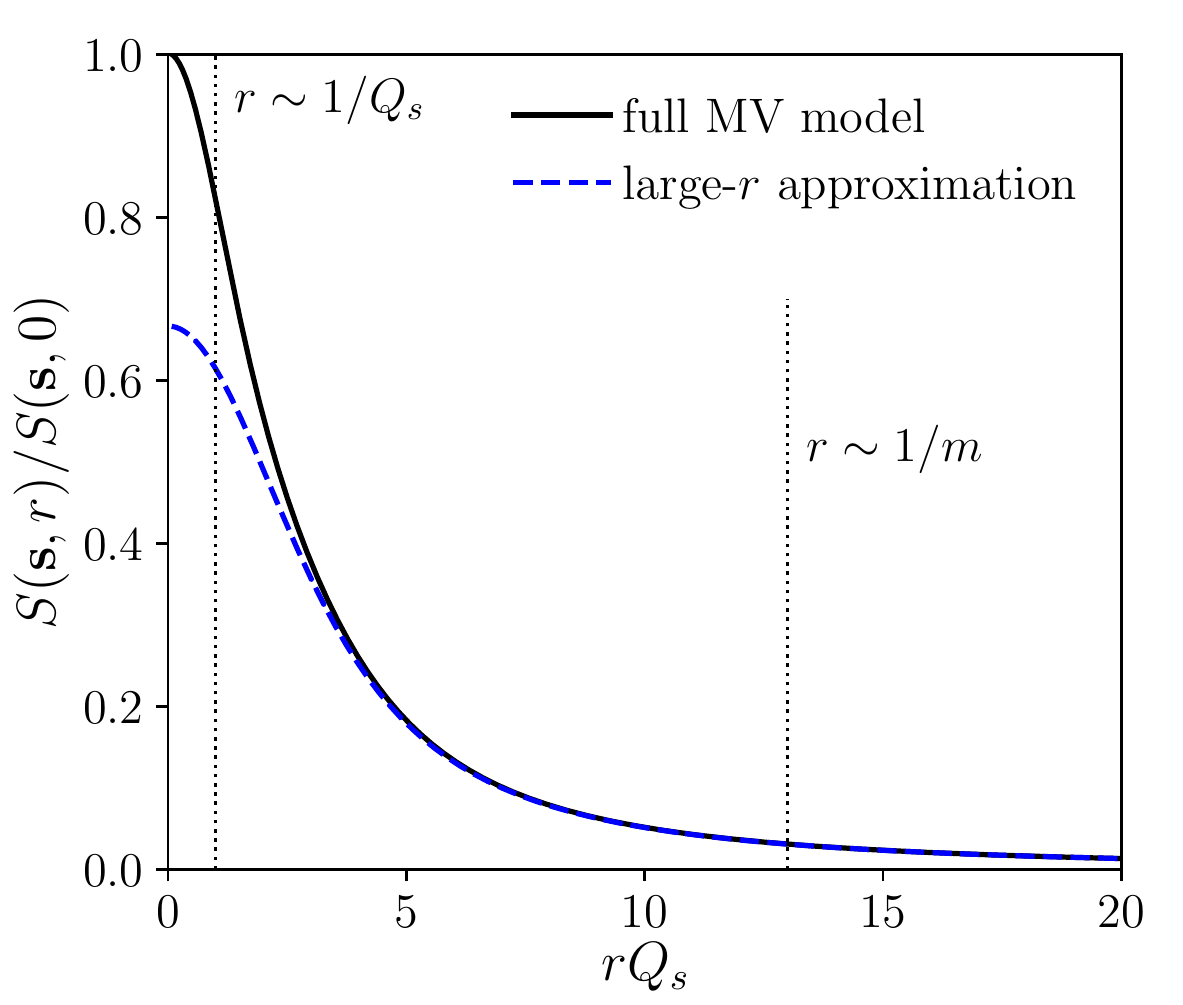}
    \caption{The solid line displays $S({\bf s},r)$ (normalized to 1 at the origin), obtained by adding the Eqs.~(4.48) and (4.49) of~\cite{Albacete:2018bbv} with $N_c=3$, in the symmetric case where the saturation scales are identical for both nuclei, as function of the dimensionless variable $rQ_s$.
      The dashed line is the large-distance contribution defined by Eq.~(\ref{eq:approx}).}
    \label{fig:5}
\end{figure}
We shall work within an extended version of the McLerran-Venugopalan (MV) model~\cite{McLerran:1993ni}, characterized by an explicit dependence on transverse coordinates introduced in the 2-point correlator of the color charge fluctuations in a large nucleus:
\begin{align}
\label{eq:mv}
\langle \rho^{a}(s_1^-,{\bf s}_1)&\rho^{b}(s_2^-,{\bf s}_2)\rangle = \\ = \nonumber &\delta^{ab}\mu^2(s_1^-)\delta(s_1^--s_2^-)h({\bf s})\delta^{(2)}({\bf r}),
\end{align}
where ${\bf s}$ and ${\bf r}$ were introduced right before Eq.~(\ref{eq:2p}), $h({\bf s})$ is the transverse profile of a nucleus, and $\mu^2(s^-)$ is the number density of color sources along the longitudinal coordinate.

The authors of Ref.~\cite{Albacete:2018bbv} derived, using this model, the connected 2-point function of the energy-density of the Glasma, i.e., the system created immediately after two large boosted nuclei cross each other.\footnote{We shall take the expressions of Ref.~\cite{Albacete:2018bbv} in the Golec-Biernat--W\"usthoff (GBW) model, thus neglecting any
$\alpha_s \ln(Q/m)$ dependence of the saturation scales, being Q either a UV cutoff or $1/r$.}
They showed that it is essentially identical to the sum of the two first terms of its $N_c$-expansion, which are given by Eqs.~(4.48) and (4.49) of Ref.~\cite{Albacete:2018bbv}.
This correlator is displayed in Fig.~\ref{fig:5} as a function of the dimensionless variable $rQ_s$. 
It is very sharp, as it decreases by one order of magnitude around $rQ_s \sim 5$, which corresponds to a length scale typically smaller than the size of a nucleon.

We now justify Eq.~(\ref{eq:2p}).
The left-hand side of Eq.~(\ref{eq:2p}) is an integral over $r$ with an upper cutoff at $1/m$:
\begin{align}
\label{eq:xis}
  \xi({\bf s})\equiv
  \int_{0}^{1/m} 2\pi r dr S( {\bf s}, r ).
\end{align}
Instead of using the full expression of $ S( {\bf s}, r )$, we identify the leading contribution to $\xi({\bf s})$.
The correlator $S( {\bf s}, r )$ falls off like $1/r^2$ for large $r$, which generates a logarithmic divergence of the integral (\ref{eq:xis}), which is regulated by the infrared cutoff. 
Hence, despite the rapid fall-off seen in Fig.~(\ref{fig:5}), the integral is actually {\it dominated\/} by the contribution at large $r$.
Following Eqs.~(4.48) and (4.49) of Ref.~\cite{Albacete:2018bbv} in the GBW limit, we extract the leading contribution for large $r$, which we denote by $S_\infty( {\bf s}, r)$. We obtain:
\begin{align}
  \label{eq:approx}
  \nonumber S_{\infty}&( {\bf s}, r )\equiv \frac{(N_c^2-1)}{2g^4N_c^2} Q_A^4Q_B^4\times\\
& \times \left[\frac{Q_B^2 r^2/4-1+e^{-Q_B^2r^2/4}}{(Q_B^2 r^2/4)^2} +(A\leftrightarrow B)\right].
\end{align}
This contribution is displayed as a dashed line in Fig.~\ref{fig:5}.
It rapidly converges to the full result for large $r$. 
Replacing $S( {\bf s}, r )$ with $S_\infty({\bf s},r)$ in Eq.~(\ref{eq:xis}) and carrying out the integral, one obtains
\begin{align}
 \nonumber \xi({\bf s})&= \frac{2\pi(N_c^2-1)}{g^4N_c^2}\times \\
  &\times \left[  Q_{A}^4Q_{B}^2
    \left(\ln\left(\frac{Q_B^2}{4 m^2}\right)-1+\gamma\right)+(A\leftrightarrow B)\right],
\end{align}
where $\gamma$ is Euler's constant. 
Eq.~(\ref{eq:2p}) is obtained by keeping the leading logarithm in this expression, and neglecting the constant term.
This leading logarithmic approximation appears to be a robust feature of the CGC description, in the sense that it would have the same form even if one relaxes the MV assumptions leading to Eq.~(\ref{eq:mv}).
Note that in Eq.~(\ref{eq:2p}) we replace $\ln x$ with $\ln(1+x)$ to ensure that $\xi({\bf s})$ is positive, even in regions where the hierarchy $Q_A,Q_B\gg m$ does not hold.
The sensitivity of our results to this regulator is studied in Appendix~\ref{sec:b}. 

\begin{figure}[t!]
    \centering
    \includegraphics[width=\linewidth]{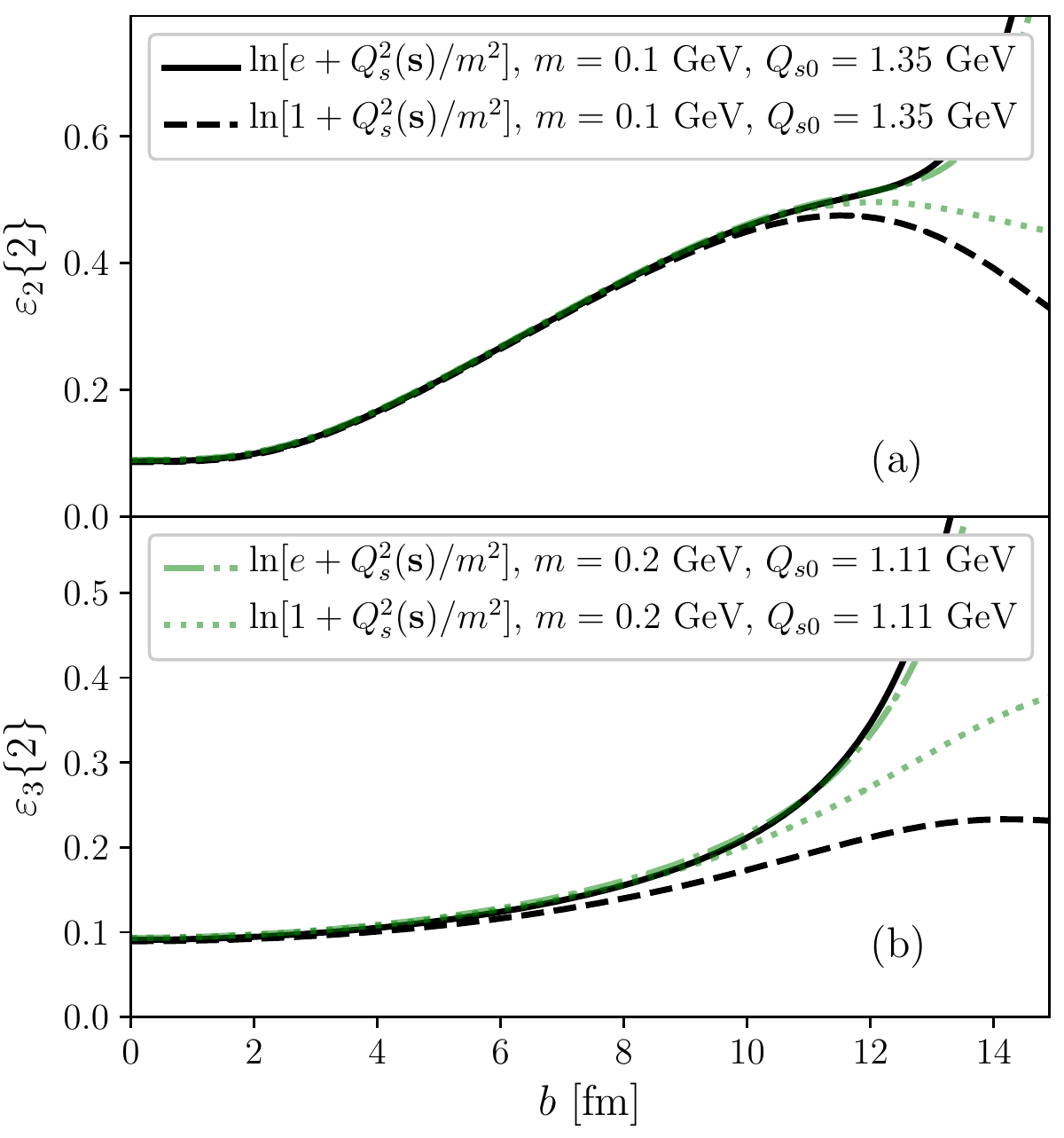}
    \caption{The figures shows the rms values of $\varepsilon_2$ [panel(a)] and $\varepsilon_3$ [panel (b)] as function of impact parameter, calculated from Eqs.~(\ref{eq:epBfinal}). Different line styles represent different combinations of the parameter $m$ and $Q_0$, or different prescriptions used to regulate the logarithms in Eq.~(\ref{eq:2p}).}
    \label{fig:6}
\end{figure}

\section{Assessing the robustness of the results}
\label{sec:b}

Our calculations assume that there is a large separation of scales between $Q_s$ and $m$, as stated by Eq.~(\ref{eq:separation}). 
However, when evaluating Eqs.~(\ref{eq:epBfinal}), we are effectively integrating over regions at the periphery of the nuclei where this ordering breaks down.
The ordering also breaks down for large impact parameters. 
The resulting errors can be evaluated by comparing different prescriptions for regulating the logarithms in Eqs.~(\ref{eq:2p}).
When $Q_{s0}\sim m$, the logarithms start to depend strongly on the specific regulator used (e.g., +1 in Eq.~(\ref{eq:2p})).
Any visible dependence of the results under variation of this regulator will indicate the break down of our formalism.

Second, we check that physical results are independent of the infrared cutoff $m$.
As explained in Sec.~\ref{sec:4}, when one varies $m$, one should renormalize $Q_{s0}$ in such a way that the right-hand side of Eq.~(\ref{eq:control}) is unchanged.

In Fig.~\ref{fig:6}, we perform an explicit check of the robustness of our results on anisotropy fluctuations under the previous conditions.
We compute $\varepsilon_2\{2\}$ and $\varepsilon_3\{2\}$ as function of impact parameter for different combinations of $m$ and $Q_{s0}$, and different prescriptions to regulate the logarithms in Eq.~(\ref{eq:2p}).
One sees that results are independent of the regulator and of the infrared cutoff except for peripheral collisions, as expected.
The error on $Q_{s0}$, inferred from its dependence on the infrared cutoff $m$, is of order $0.2$~GeV.

\end{document}